\documentclass[12pt,3p]{elsarticle}
\biboptions{sort&compress}

\usepackage[utf8]{inputenc}
\usepackage[T2A,T1]{fontenc}
\usepackage[russian,english]{babel}

\usepackage{amsmath,amsfonts,amssymb,amsthm}
\usepackage{bm}
\usepackage {graphicx}
\graphicspath{{fig/}}

\usepackage{xcolor}

 \usepackage{hyperref}
\hypersetup{
	colorlinks=true,        
	linkcolor=blue,          
	citecolor=blue,        
	filecolor=magenta,      
}

\usepackage{array}
\usepackage{comment}

\allowdisplaybreaks[4]

\usepackage{color} 				
\definecolor{dgray}{rgb}{0.6,0.6,0.6}
\definecolor{dmag}{rgb}{0.6,0.0,0.6}
\definecolor{mbul}{rgb}{0.102, 0.42, 0.102} 
\usepackage[normalem]{ulem} 	

\definecolor{pink}{rgb}{1,0,0.9}

\newcommand{\pd}{{\phantom{\dagger}}}

\newcommand{\Tr}{{\rm Tr\,}}

\begin{document}

\title{General collisionless kinetic approach to studying excitations in arbitrary-spin quantum atomic gases}



\author[1,2]{M.~Bulakhov}
\ead{bulakh@kipt.kharkov.ua}
\author[1,2]{A.S.~Peletminskii}
\ead{aspelet@kipt.kharkov.ua}
\author[1,2,3]{Yu.V.~Slyusarenko}
\affiliation[1]{
orgsnization={Akhiezer Institute for Theoretical Physics, National Science Center "Kharkiv Institute of Physics and Technology", NAS of Ukraine}, 
city={Kharkiv}, 
postcode={61108},  
country={Ukraine}}
\affiliation[2]{
organization={V.N. Karazin Kharkiv National University},
city={Kharkiv},
postcode={61022},
country={Ukraine}
}
\affiliation[3]{
    organization={Lviv Polytechnic National University},
    city={Lviv},
    postcode={79000}, 
    country={Ukraine}
}




\begin{abstract}
We develop a general kinetic approach to studying high-frequency collective excitations in arbitrary-spin quantum gases. To this end, we formulate a many-body Hamiltonian that includes the multipolar exchange interaction as well as the coupling of a multipolar moment with an external field.  By linearizing the respective collisionless kinetic equation, we find a general dispersion equation that allows us to examine the high-frequency collective modes for arbitrary-spin atoms obeying one or another quantum statistics. We analyze some of its particular solutions describing spin waves and zero sound for Bose and Fermi gases.    
\end{abstract}


\maketitle

\section{Introduction}

Over the last three decades, ultracold gases loaded in magneto-optical traps have become an ideal platform for studying quantum collective phenomena that are difficult to probe in realistic materials \cite{Lewenstein_AdvPhys_2007,Bloch_RMP_2008,PitStr2016,Pethick2008}. 
This is due to the fact that such systems are extremely pure and many physical parameters like density, temperature, spin, dimensionality and even interaction strength are in a high degree of experimental control.

The kinetic approach, which is based on the assumptions of weak interatomic interaction and inhomogeneity of the system, is a powerful tool for describing and setting up experiments. It allows one to examine the collective modes in dilute systems such as quantum gases \cite{Klimontovich_Dokl_1952, klimontovich_JETP_1952} and plasma \cite{Akhiezer1975}. The excitation spectrum represents the important characteristics since it determines the physical properties of the system and its stability. Therefore, a priority problem is to find the possible branches of the excitation spectrum and to identify the conditions for a given part of the spectrum to be main and determining one.   

The so-called high-spin ($F>1/2$) dilute atomic gases show a rich variety of excitations and other collective phenomena. This is also due to the fact that quantum gases exist in a superfluid state. In particular, even the simplest high-spin system as a weakly interacting gas of spin-1 atoms with Bose-Einstein condensate demonstrates the emergence of ferromagnetic, polar (quadrupolar), and paramagnetic states with respective branches of single-particle excitations \cite{Ohmi_JPSJ_1998,Ho_PRL_1998,Akhiezer_JETP_1998,Peletminskii_PhysA_2007}. The fourth magnetic state, known as the broken-axisymmetry phase \cite{Stenger_Nat_1998,Ueda_PRA_2007,Dalibard_PRA_2012,Bulakhov_JPhysA_2022}, occurs due to the quadratic Zeeman effect, which reflects the coupling of the magnetic field with the spin-quadrupolar moment, an attribute of the high-spin system (for details, see reviews \cite{Ueda_PhysRep_2012,Ueda_RMP_2013} on spinor Bose gases). The normal-state spin-1 gas has also a rich array of spin excitations \cite{Endo_JLTP_2008,Natu_PRA_2010}. As for gases of high-spin fermionic atoms, they exhibit a wealth of novel collective effects: from zero sound \cite{Yip_PRA_1999,Bulakhov_JPhysA_2023} and spin waves \cite{Lewenstein_PRL_2013} to a variety of superfluid phenomena caused by high-spin Cooper pairs \cite{Ho_PRL_1999}.   

In contrast to $F=1/2$ systems, whose description requires to employ the identity and three spin operators, high-spin systems should be consistently studied by involving additional multipolar operators, which are organized into higher-order tensors (quadrupolar, octupolar, etc.). The role of multipolar operators in the Hamiltonian is twofold: they are responsible for the coupling of the multipolar moment with the applied magnetic field and for the interatomic exchange interaction. 
In particular, the experimentally relevant inhomogeneous trap field is described by two effective constant parameters \cite{Stenger_Nat_1998} associated with the gradient field (linear Zeeman effect) and bias field (quadratic Zeeman effect). These parameters can be related to the components of the reducible two-rank tensor that specify the external field \cite{Bulakhov_JPhysA_2022}. In turn, the components themselves are coupled with multipolar operators. Regarding the interatomic interaction, in general case of non-contact potential, there is an exchange of any multipolar moment \cite{Bulakhov_JPhysA_2023}, whereas for a contact potential parameterized by $s$-wave scattering length, the number of moments involved in the exchange multipolar interaction is significantly reduced \cite{Yip_PRA_1999}. 

For a particular spin-1 system, the mentioned multipolar operators can be realized by means of the Gell-Mann generators of the SU(3) Lie algebra \cite{Peletminskii_PLA_2020,Bulakhov_JPhysA_2021} or by other representations \cite{Kiselev_EPL_2013}. Stevens operators \cite{Stevens_Proceed_1952} are also employed to introduce them (for spin-3/2 system, see \cite{Kosmachev_JETP_2015}). However, the most elegant and straightforward way to treat multipolar operators and associated high-spin physics is to use the apparatus of irreducible spherical tensor operators \cite{Wigner_1959,Racah_1959,Brink_1968,Sakurai_2020}. In the context of spin-3/2 quantum gases, the latter was applied while studying collective excitations such as spin waves \cite{Lewenstein_PRL_2013} and zero sound \cite{Bulakhov_JPhysA_2023}.

Although there are already a number of works on collective excitations in high-spin gases, we aim to give an elegant kinetic formulation through for arbitrary spin the apparatus of spherical tensor operators  (both quantum statistics). This formulation includes both multipolar exchange interaction of non-contact type and external field tensor whose rank limited by the total atomic spin. To this end, in Section \ref{sec:Hamilt}, we construct and justify the general many-body Hamiltonian. It includes two ingredients associated with high atomic spin: the coupling of any multipolar moment with external field and existence of multipolar exchange interaction. Both of them are treated through the irreducible spherical tensor operators. Then we present the respective collisionless kinetic equation valid for bosonic and fermionic atoms. Its linearization performed in Section \ref{sec:Linearized} gives the general dispersion equation which allows one to explore the high-frequency collective excitations. In Section \ref{sec:Solution}, we analyze some of its particular solutions for deeply degenerate Fermi and Bose gases. Finally, we discuss the obtained results in Section \ref{sec:Discussion}. 

\section{Hamiltonian and kinetic equation for arbitrary-spin atoms}\label{sec:Hamilt}

Let us consider the Hamiltonian describing a many-particle system of high-spin identical atoms with pairwise interaction $H_{\rm int}$, in an external field. A specific feature of such system is that its Hamiltonian, in the general case, should be constructed not only from spin operators but also from multipolar moment operators. The most elegant way to include all these degrees of freedom into the Hamiltonian is to employ the apparatus of irreducible spherical tensor operators \cite{Wigner_1959,Racah_1959,Brink_1968,Sakurai_2020}. For spin-$F$ atoms, the respective rotationally-invariant total Hamiltonian can be written as
\begin{equation} \label{eq:TotHam}
H=H_{0}+H_{\rm int},
\end{equation}
where 
\begin{equation}     
    H_{0}
    =
    \sum_{{\bf p}}
    a^{\dagger}_{{\bf p}\alpha}
    \left[
        \varepsilon_{\bf p}
        \delta_{\alpha\beta}
        -
        \sum_{j=0}^{2F}
        \sum_{m=-j}^{j}(-1)^{m}h^{j}_{m}(T^{j}_{-m})_{\alpha\beta}
    \right]
    a_{{\bf p}\beta}
    ,
    \label{eq:CouplHam}
\end{equation}
$a^{\dagger}_{{\bf p}\alpha}$ and $a_{{\bf p}\alpha}$ are the creation and annihilation operators of bosons or fermions depending on the integer or half-integer value of the total atomic spin $F$. The operator
$H_0$ includes the kinetic energy term $\varepsilon_{\bf p}=p^{2}/2M$, as well as the coupling of any multipolar moment with the external field. This coupling is specified by two irreducible tensors $T^{j}_{m}$ and $h^{j}_{m}$ with indices $j$ and $m$ denoting their rank and component, respectively (for a given rank $j$, both tensors have $2j+1$ components). The first one, being a spherical tensor operator $T^{j}_{m}$, describes the multipolar degrees of freedom. Among the components of the second tensor $h^{j}_{m}$ there are those that reflect the linear and quadratic Zeeman effects, so that:
\begin{equation}
    h^{1}_{0}T^{1}_{0}
    \equiv 
    g\mu_{\rm B}H_{z}F_{z}, 
    \quad
    h^{2}_{0}T^{2}_{0}
    \equiv
    \frac{(g\mu_{B}H_{z})^{2}}{E_{\rm m}-E_{\rm i}}F_{z}^{2}+q_{\mathrm{MW}}F_{z}^{2},
\end{equation}
where $g$ is the Lande factor, $\mu_{\rm B}$ is the Bohr magneton, $H_{z}$ is the magnetic field directed along $z$-axis, $F_{z}$ is $z$-component of the spin operator, $E_{\mathrm{m}}-E_{\mathrm{i}}$ is the hyperfine energy splitting given by the initial ($E_{\mathrm {i}}$) and intermediate ($E_{\mathrm{m}}$) energies. The contribution $q_{\mathrm{MW}}$ to the quadratic Zeeman coupling is experimentally produced by a microwave \cite{Gerbier2006,Leslie2009} or light \cite{Santos2007} field, so that $h^{1}_{0}$ and $h^{2}_{0}$ can be varied arbitrarily \cite{Stenger_Nat_1998}. The rest of tensorial components $h^{j}_{m}$ are included into the Hamiltonian for the sake of generality and symmetry of formulation. Thus, the Hamiltonian \eqref{eq:CouplHam} describes all physical effects induced by the spin of an atom.  

As for the isotropic interaction Hamiltonian $H_{\rm int}$ in Eq.~(\ref{eq:TotHam}), it can be written in the following form: \cite{Bulakhov_JPhysA_2023}:
\begin{flalign} 
    &H_{\rm int}
    =
    \nonumber\\
    &\frac1{2 V}
    \sum_{j=0}^{2F}
    \sum_{m=-j}^{j}
    (-1)^{m}
    \sum_{{\bf p}_{1}\ldots{\bf p}_{4}}
    U^{[j]}({\bf p}_{1}-{\bf p}_{4})a^{\dagger}_{{\bf p}_{1}\alpha}a^{\dagger}_{{\bf p}_{2}\beta}
    (T^{j}_{m})_{\alpha\delta}(T^{j}_{-m})_{\beta\gamma}
    a^\pd_{{\bf p}_{3}\gamma}
    a^\pd_{{\bf p}_{4}\delta}\,\delta^\pd_{{\bf p}_{1}+{\bf p}_{2},\,{\bf p}_{3}+{\bf p}_{4}},
\label{eq:InterHam}
\end{flalign}
where ${V}$ is the volume of the system and $U^{[j]}({\bf p}_{1}-{\bf p}_{4})$ are the Fourier transforms of the energies
corresponding to the direct ($j=0$) interaction, as well as the multipolar ($j=1,\dots, 2F$) exchange interactions. Note that when the interaction is parameterized by the $s$-wave scattering length (contact potential), the multipolar exchange does not include all the multipolar moments \cite{Yip_PRA_1999,Peletminskii_PLA_2020,Bulakhov_JPhysA_2021,Bulakhov_JPhysA_2022,Bulakhov_JPhysA_2023}.
In equations (\ref{eq:CouplHam}), (\ref{eq:InterHam}) and below, we assume summation over the repeated Greek indices associated with the spin projection, unless otherwise specified. In addition, indices in square brackets do not imply summation. 

The next step is to derive the relevant kinetic equation for the Hamiltonian given by Eqs.~\eqref{eq:TotHam}--\eqref{eq:InterHam}. In general case, this can be rigorously done by employing the reduced description method of quantum many-body systems \cite{Yatsenko1968, AkhPel}. 
The main ingredient of this method is a coarse-grained statistical operator which depends on time through a number of so-called reduced description parameters (master variables). For quantum gases in the normal state, the single-particle density matrix $f_{\alpha\beta}(\mathbf{p},\mathbf{p}')$ can be chosen as such parameter. From a perturbative expansion of the corresponding coarse-grained statistical operator, one can obtain the kinetic equation for the single-particle density matrix with the collision integral in the second and higher orders in interaction \cite{AkhPel}. In terms of the Wigner density matrix,
\begin{equation*}
       f_{\alpha\beta}
    ({\bf x},{\bf p})
    =\sum_{\bf k}e^{-i{\bf kx}}f_{\alpha\beta}\left({\bf p}-\frac{{\bf k}}{2},{\bf p}+\frac{{\bf k}}{2}\right),
\end{equation*}
which describes states with weak inhomogeneity (inhomogeneity of small amplitude and large wavelength), this kinetic equation takes the following form \cite{Bulakhov_JPhysA_2023}: 
\begin{gather}
    \partial_t
    f_{\alpha\beta}
    (t,{\bf x},{\bf p})
    +
    \frac{i}{\hbar}
    \left[
        \varepsilon
        (t,{\bf x},{\bf p})
        ,
        f
        (t,{\bf x},{\bf p})
    \right]
    _{\alpha\beta}
    \nonumber\\+
    \frac12
    \left\{
        \partial_{\mathbf{p}}
        \varepsilon(t,{\bf x},{\bf p})
        ,
        \partial_{\mathbf{x}}
        f(t,{\bf x},{\bf p})
    \right\}_{\alpha\beta}
    -
    \frac12
    \left\{
        \partial_{\mathbf{x}}
        \varepsilon(t,{\bf x},{\bf p})
        ,
        \partial_{\mathbf{p}}
        f(t,{\bf x},{\bf p})
    \right\}_{\alpha\beta}
    =
    L^{(2)}[f],
    \label{eq:KinEqX}
\end{gather}
with the particle mean-field energy $\varepsilon_{\alpha\beta}({\bf x},{\bf p})$ defined by 
\begin{gather}
    \varepsilon_{\alpha\beta}(t,{\bf x},{\bf p})
    =
    \varepsilon_{\bf p}\delta_{\alpha\beta}
    -
    \sum_{j=0}^{2F}
    \sum_{m=-j}^{j}(-1)^{m}
    h^{j}_{m}
    (T^{j}_{-m})_{\alpha\beta} 
    \nonumber  
    \\
    +
    \frac{1}{V}
    \sum_{j=0}^{2F}
    \sum_{m=-j}^{j}
    (-1)^{m}
    \sum_{{\bf p}'}
    \int d^3x'\,
    \mathcal{U}^{[j]}({\bf x}-{\bf x}')
    (T^{j}_{m})_{\alpha\beta}
    (T^{j}_{-m})_{\delta\gamma}f_{\gamma\delta}(t,{\bf x}',{\bf p}') 
    \nonumber \\
    \pm 
    \frac{1}{V}
    \sum_{j=0}^{2F}
    \sum_{m=-j}^{j}(-1)^{m}
    \sum_{{\bf p}'}
    U^{[j]}({\bf p}-{\bf p}')
    (T^{j}_{m})_{\alpha\gamma}
    f_{\gamma\delta}(t,{\bf x},{\bf p}')
    (T^{j}_{-m})_{\delta\beta}
    , \label{eq:QuasEnX}
\end{gather}
where 
\begin{equation*}
    U^{[j]}({\bf p})=\int\mathop{d^3x}\mathcal{U}^{[j]}({\bf x})e^{-\tfrac{i}{\hbar}{\bf px}}. 
\end{equation*}
For weakly inhomogeneous states, $f_{\alpha\beta}(\mathbf{p}-{\mathbf k}/2,\mathbf{p}+{\mathbf k}/2)$ has a sharp maximum at $\mathbf{k}=0$ so that $f_{\alpha\beta}(\mathbf{p}-{\mathbf k}/2,\mathbf{p}+{\mathbf k}/2)=f(\mathbf p)\delta_{\mathbf{k}0}$ for the homogeneous case. 
The collision integral $L^{(2)}[f]$ is of the second order in interaction. Therefore, for weak interactions, it can be neglected when considering processes that are determined by the first order mean-field terms. 
In other words, the collision integral can be estimated as $L^{(2)}[f]\propto\delta f_{\alpha\beta}/\tau$, where $\delta f_{\alpha\beta}$ is the deviation of the density matrix from its equilibrium value and $\tau$ is the relaxation time. Consequently, it can be omitted if we are interested in high-frequency collective modes, $\omega\tau\gg 1$, and do not study their damping due to collisions. In atomic gases with internal degrees of freedom, there may be several relaxation times. The shortest time is due to direct collisions. 
Therefore, for gases with weak interactions, all relaxation times are large. 
The left-hand side of Eq.~\eqref{eq:KinEqX} has the same form both for fermions and bosons. Quantum statistics affects only the explicit form of the collision integral \cite{AkhPel} and the expression for the mean-field particle energy. As for the latter, the sign minus (plus) in the last term of Eqs.~\eqref{eq:QuasEnX} corresponds to the Fermi (Bose) statistics. 

As for Eq.~\eqref{eq:QuasEnX}, it can be essentially simplified by noting that the interaction potential $\mathcal{U}^{[j]}({\bf x}-{\bf x}')$ has a sharp peak at $\mathbf{x}=\mathbf{x}'$, whereas the Wigner density matrix experiences small spatial variations (weak inhomogeneity). Thus, in the third term, the distribution function $f_{\gamma\delta}({\bf x}',{\bf p}')$ can be taken out of the integral at the point ${\bf x}$ and Eq.~ \eqref{eq:QuasEnX} is reduced to 
\begin{gather}
    \varepsilon_{\alpha\beta}({\bf x},{\bf p})=\varepsilon_{\bf p}\delta_{\alpha\beta}
    -
    \sum_{j=0}^{2F}
    \sum_{m=-j}^{j}(-1)^{m}h^{j}_{m}(T^{j}_{-m})_{\alpha\beta}
    \nonumber\\
    +\frac{1}{V}\sum_{j=0}^{2F}\sum_{m=-j}^{j}(-1)^{m}\sum_{{\bf p}'} U^{[j]}(0)(T^{j}_{m})_{\alpha\beta}(T^{j}_{-m})_{\delta\gamma}f_{\gamma\delta}({\bf x},{\bf p}') \nonumber \\
\pm\frac{1}{V}\sum_{j=0}^{2F}\sum_{m=-j}^{j}(-1)^{m}\sum_{{\bf p}'}U^{[j]}({\bf p}-{\bf p}')(T^{j}_{m})_{\alpha\gamma}f_{\gamma\delta}({\bf x},{\bf p}')(T^{j}_{-m})_{\delta\beta}.\label{eq:QuasEnXSimp}
\end{gather}
It is worth noting that the obtained kinetic equation \eqref{eq:KinEqX} is valid if the characteristic scale of spatial inhomogeneity (the distance over which the distribution function changes) is large compared to the interaction range and the de Broglie wavelength. This equation is applicable to study both homogeneous and inhomogeneous quantum gases in the normal state. For superfluid gases with broken U(1) symmetry, the number of the reduced description variables enlarges due to the occurrence of order parameters associated with anomalous averages. The pair anomalous averages, responsible for Cooper pairing in BCS superfluidity, can also be important for Bose systems with condensates \cite{Stoof_PRA_1994,Peletminskii_LTP_2010, Griffin_PRB_1996,Giorgini_PRL_1998,Poluektov_CMP_2013}. 
Therefore, the derivation of the respective kinetic equations should be extensively modified both for Bose \cite{Shchelokov_TMP_1977,Kirkpatrick_JLTP_1985,Walser_PRA_1999,Reichl_PRA_2013,Reichl_JPhysA_2019,Tran_PRE_2020} and Fermi gases \cite{Galaiko_JETP_1972,Krasil'nikov_PhysA_1990}.  

Next, we decompose the Wigner density matrix and the particle energy into a complete set of irreducible spherical tensor operators:
\begin{gather} 
    f_{\alpha\beta}(t,{\bf x},{\bf p})=
    \sum_{j=0}^{2F}\sum_{m=-j}^{j}
    (-1)^{m}f^{j}_{m}(t,{\bf x},{\bf p})
    (T^{j}_{-m})_{\alpha\beta}
    , \nonumber \\
    \varepsilon_{\alpha\beta}(t,{\bf x},{\bf p})=
    \sum_{j=0}^{2F}\sum_{m=-j}^{j}
    (-1)^{m}
    \varepsilon^{j}_{m}(t,{\bf x},{\bf p})
    (T^{j}_{-m})_{\alpha\beta}
    .\label{eq:Decomp}
\end{gather}
Then by using the normalization condition given by Eq.~\eqref{eq:Norm}, the respective coefficients are found to be,
\begin{gather} 
    f^{j}_{m}(t,{\bf x},{\bf p})
    =
    f_{\alpha\beta}(t,{\bf x},{\bf p})
    (T^{j}_{m})_{\beta\alpha}
    ,\nonumber
    \\
    \varepsilon^{j}_{m}(t,{\bf x},{\bf p})
    =
    \varepsilon_{\alpha\beta}(t,{\bf x},{\bf p})(T^{j}_{m})_{\beta\alpha}
    .
    \label{eq:DecCoeff}
\end{gather}
Note that it is the coefficients $f^{j}_{m}$ that determine, in the spherical basis, the physical quantities such as three components of the magnetization vector for $j=1$, five components of the quadrupolar tensor for $j=2$, seven components of the octupolar tensor for $j=3$, etc. The kinetic equation in their terms can be written in the following form:
\begin{gather}
    \partial_t 
    f_{m}^{j}(t,{\bf x},{\bf p})
    +
    \frac{i}{\hbar}
    B_{m;m_{1}m_{2}}^{\,j;\,j_{1}\,j_{2}}
    \varepsilon_{m_{1}}^{j_{1}}(t,{\bf x},{\bf p})
    f_{m_{2}}^{j_{2}}(t,{\bf x},{\bf p})
    \nonumber \\
    +
    \frac{1}{2}C_{m;m_{1}m_{2}}^{\,j;\,j_{1}j_{2}}
    \left(
        \partial_{\bf p}
        \varepsilon_{m_{1}}^{j_{1}}(t,{\bf x},{\bf p})
        \partial_{\mathbf{x}} 
        f_{m_{2}}^{j_{2}}(t,{\bf x},{\bf p})
        -
        \partial_{\mathbf{x}}
        \varepsilon_{m_{1}}^{j_{1}}(t,{\bf x},{\bf p})
        \partial_{\bf p} f_{m_{2}}^{j_{2}}(t,{\bf x},{\bf p})
    \right)
    =
    L^{(2)}[f]
    ,
    \label{eq:KinEquat}
\end{gather}
where    
\begin{gather}
B_{m;m_{1}m_{2}}^{j;j_{1}j_{2}}=(-1)^{m_1+m_2}\,{\rm Tr}\,\left(T^{j}_{m}\left[T_{-m_{1}}^{j_{1}},T_{-m_{2}}^{j_{2}}\right]\right), \nonumber \\
C_{m;m_{1}m_{2}}^{j;j_{1}j_{2}}=(-1)^{m_1+m_2}\,{\rm Tr}\left(\,T^{j}_{m}\left\{T_{-m_{1}}^{j_{1}},T_{-m_{2}}^{j_{2}}\right\}\right). \label{eq:BCCoeff}
\end{gather}
These coefficients have the following evident symmetry properties:
\begin{gather}
B_{m;m_{1}m_{2}}^{j;j_{1}j_{2}}=B_{m_{1}m_{2};m}^{j_{1}j_{2};j}=-B_{m;m_{2}m_{1}}^{j;j_{2}j_{1}}, \quad B^{j;j_{1}j_{2}}_{m;0\,0}=0, \nonumber \\
B^{0;j_{1}j_{2}}_{0;m_{1}m_{2}}=B^{j;0j_{2}}_{m;0\,m_{2}}=B^{j;j_{1}0}_{m;m_{1}0}=0, \nonumber \\
C_{m;m_{1}m_{2}}^{j;j_{1}j_{2}}=C_{m_{1}m_{2};m}^{j_{1}j_{2};j}=C_{m;m_{2}m_{1}}^{j;j_{2}j_{1}}. \label{eq:PropBC}
\end{gather}
Moreover, due to the selection rule \eqref{eq:SelRule} for spherical tensor operators, we have 
\begin{equation}\label{eq:SelRuleBC}
    B_{m;m_{1}m_{2}}^{j;j_{1}j_{2}}=0, \quad  C_{m;m_{1}m_{2}}^{j;j_{1}j_{2}}=0, \quad {\rm if} \quad  m-m_{1}-m_{2}\neq0.
\end{equation}
Finally, substituting \eqref{eq:QuasEnXSimp} into \eqref{eq:DecCoeff} and employing the Friez-like identity \eqref{eq:Fierz-like}, one obtains the particle energy which accounts for the mean-field effects and governs the kinetic equation \eqref{eq:KinEquat},  
\begin{gather}
    \varepsilon_{m}^{j}(t,{\bf x},{\bf p})
    =
    \sqrt{2F+1}
    \varepsilon_{\bf p}
    \delta_{j0}\delta_{m0}
    -
    h^{j}_{m}
    +
    \frac1V
    \sum_{{\bf p}'}
    J^{[j]}({\bf p}-{\bf p}')
    f^{j}_{m}(t,{\bf x},{\bf p}'),
\nonumber \\
    J^{[j]}({\bf p}-{\bf p}')
    =
    U^{[j]}(0)
    \pm
    \sum_{j'=0}^{2F}
    U^{[j']}({\bf p}-{\bf p}')A^{j'j}.
    \label{eq:PartEn}
\end{gather}
Here matrix $A^{ij}$, given by \eqref{eq:Fierz-like}, mixes the multipolar degrees of freedom determined by $f^{j}_{m}({\bf x},{\bf p}')$ with their exchange interaction energies.

\section{Linearized kinetic equation}\label{sec:Linearized}

Now we proceed to solving the formulated kinetic equation \eqref{eq:KinEquat} for a given particle energy \eqref{eq:PartEn}. In doing so, we restrict ourselves to its approximate solution assuming that the tensorial components of the Wigner distribution function $f^{j}_{m}({\bf x},{\bf p})$ slightly deviate from a homogeneous stationary state,
\begin{equation}
    f^{j}_{m}(t,{\bf x},{\bf p})=
    (f^{j}_{0})_{\mathbf{p}}\delta_{m0} 
    +
    (\tilde{f}^{j}_{m})_{\mathbf{p}} 
    ,
    \label{eq:WignDev}
\end{equation}
 where $(f^{j}_{0})_{\mathbf{p}}\equiv f^{j}_{0}({\mathbf{p}})$ is the equilibrium distribution function and $(\tilde{f}^{j}_{m})_{\mathbf{p}}\equiv\tilde{f}^{j}_{m}(t,{\mathbf{x}},{\mathbf{p}})$ represents a perturbation. This approximate solution describes the weakly excited states of the system under consideration. Since the equilibrium density matrix is diagonal $f_{\alpha\beta}(\mathbf{p})=f^{[\alpha]}({\bf p})\delta_{\alpha\beta}$ and only $T^{j}_{0}$ by its definition have non-zero diagonal elements (see \ref{app:STO}), one finds from Eq.~\eqref{eq:DecCoeff},
 \begin{equation}
      (f^{j}_{0})_{\mathbf{p}}=f^{[\alpha]}({\bf p})(T^{j}_{0})_{\alpha\alpha}
      .
      \label{eq:EqWig}
 \end{equation}
Equation \eqref{eq:WignDev}, according to \eqref{eq:PartEn}, also induces the deviation of the mean-field particle energy, 
\begin{equation}
    \varepsilon^{j}_{m}(t,{\bf x},{\bf p})=
    (\varepsilon^{j}_{m})_{\mathbf{p}} 
    +
    (\tilde{\varepsilon}^{j}_{m})_{\mathbf{p}}, 
    \label{eq:EnDev}
\end{equation}
where
\begin{gather}
    (\varepsilon^{j}_{m})_{\mathbf{p}} 
    =
    \sqrt{2F+1}
    \varepsilon_{\bf p}
    \delta_{j0}
    \delta_{m0}
    -h^{j}_{0}
    +
    \frac1V
    \sum_{{\bf p}'}
    J^{[j]}({\bf p}-{\bf p}')
    (f^{j}_{m})_{\mathbf{p}} 
    \delta_{m0}
    ,
    \label{eq:EquivEn}\\
    (\tilde{\varepsilon}^{j}_{m})_{\mathbf{p}} 
    =
    \frac1V
    \sum_{{\bf p}'}
    J^{[j]}({\bf p}-{\bf p}')
    (\tilde{f}^{j}_{m})_{\mathbf{p}'}. 
    \label{eq:PertEn}
\end{gather}
We assume that the external field has only a zero component caused by the existence of some axis of symmetry relevant to the experimental conditions. Performing the linearization procedure for the kinetic equation, one obtains
\begin{gather}
    \partial_{t}(\tilde{f}^{j}_{m})_{\mathbf{p}} 
    +
    \frac{i}{\hbar}B_{m;m_{1}m_{2}}^{\,j;\,j_{1}\,j_{2}}
    \left[
        (\varepsilon_{m_{1}}^{j_{1}})_{\mathbf{p}} 
        (\tilde{f}_{m_{2}}^{j_{2}})_{\mathbf{p}} 
        +
        (\tilde{\varepsilon}_{m_{1}}^{j_{1}})_{\mathbf{p}} 
        ({f}_{0}^{j_{2}})_{\mathbf{p}} \delta_{m_{2}0}
    \right]
    +
    \nonumber\\
    \frac{1}{2}C_{m;m_{1}m_{2}}^{\,j;\,j_{1}j_{2}}
    \left[
        \partial_{\mathbf{p}}
        (\varepsilon_{m_{1}}^{j_{1}})_{\mathbf{p}} 
        \partial_{\mathbf{x}}
        (\tilde{f}^{j_2}_{m_2})_{\mathbf{p}} 
        -
        \partial_{\mathbf{x}}
        (\tilde{\varepsilon}_{m_{1}}^{j_{1}})_{\mathbf{p}} 
        {\partial_{\bf p}}
        ({f}_{0}^{j_{2}})_{\mathbf{p}} 
        \delta_{m_{2}0}
    \right]
    =0
    .
\end{gather}
Let us now apply to this equation the following Fourier-Laplace transform:
\begin{equation*}
    \{\tilde{f}^{j}_{m}\}_{\mathbf{p}} 
    =
    \int d^3x
    \int_{0}^{\infty} \mathop{dt}
    e^{-i\mathbf{kx}-at}
    (\tilde{f}^{j}_{m})_{\mathbf{p}} 
    ,
\end{equation*}
where  
$\mathbf{k}$ is the wave vector, $a$ is the Laplace parameter, whose real part is assumed to be large enough for the respective integral to exist. For the sake of brevity, we omit both these parameters in all notations. Thus, the resulting kinetic equation reads,   
\begin{gather}
    a
    \{\tilde{f}^{j}_{m}\}_{\mathbf{p}}
    +
    \frac{i}{\hbar}B_{m;m_{1}m_{2}}^{\,j; ,j_{1}\,j_{2}}
    \left[
        (\varepsilon_{m_{1}}^{j_{1}})_{\mathbf{p}} 
        \{\tilde{f}_{m_{2}}^{j_{2}}\}_{\mathbf{p}}
        +
        \{\tilde{\varepsilon}_{m_{1}}^{j_{1}}\}_{\mathbf{p}}
        ({f}_{0}^{j_{2}})_{\mathbf{p}} \delta_{m_{2}0}
    \right]
    +
    \nonumber\\
    \frac{i\mathbf{k}}{2}C_{m;m_{1}m_{2}}^{\,j;\,j_{1}j_{2}}
    \left[
        \partial_{\mathbf{p}}
        (\varepsilon_{m_{1}}^{j_{1}})_{\mathbf{p}} 
        \{\tilde{f}^{j_2}_{m_2}\}_{\mathbf{p}}
        -
        \{\tilde{\varepsilon}_{m_{1}}^{j_{1}}\}_{\mathbf{p}}
        {\partial_{\bf p}}
        ({f}_{0}^{j_{2}})_{\mathbf{p}} 
        \delta_{m_{2}0}
    \right]
    =
    (g^{j}_{m})_{\mathbf{pk}}
    ,
    \label{eq:FLkineq}
\end{gather}
where $(g^{j}_{m})_{\mathbf{pk}}$ is the initial condition determined by
\begin{equation*}
    (g^{j}_{m})_{\mathbf{pk}}
    =
    \mathop{\int d^3x}
    e^{-i\mathbf{kx}}
    g^{j}_{m}(\mathbf{x},\mathbf{p})
    ,\quad
    g^{j}_{m}(\mathbf{x},\mathbf{p})
    =
    (\tilde{f}^{j}_{m})_{\mathbf{p}} \Big\vert_{t=0}.
\end{equation*}
Next, employing the fact that the unperturbed particle energy has only diagonal components ($(\varepsilon^{j}_{m})_{\mathbf{p}}\propto \delta_{m0}$, see Eq.~\eqref{eq:EquivEn}) and using the properties given by Eqs.~\eqref{eq:PropBC} and \eqref{eq:SelRuleBC}, one gets
\begin{gather}
    \sum_{j_1\geq |m|}
    \left(
        a
        \delta^{jj_1}
        +
        (G^{jj_1}_{m})_{\mathbf{p}}
    \right)
    \{\tilde{f}^{j_1}_{m}\}_{\mathbf{p}}
    =
    (g^{j}_{m})_{\mathbf{pk}}
    -
    \sum_{j_1\geq |m|}
    (R^{jj_1}_{m})_{\mathbf{p}}
    \{\tilde{\varepsilon}_{m}^{j_{1}}\}_{\mathbf{p}}
    ,
    \label{eq:Sysf}
\end{gather}
where we introduced the following quantities determined by the equilibrium values:
\begin{equation}
    (G^{jj_1}_{m})_{\mathbf{p}} 
    =
    \sum_{j_2}
    \left(
        \frac{i}{\hbar}B_{m;0m}^{\,j; ,j_{2}\,j_{1}}
        (\varepsilon_{0}^{j_{2}})_{\mathbf{p}} 
        +
        \frac{i\mathbf{k}}{2}C_{m;0m}^{\,j;\,j_{2}j_{1}}
        \partial_{\mathbf{p}}
        (\varepsilon_{0}^{j_{2}})_{\mathbf{p}} 
    \right)
    \label{eq:G}
\end{equation}
and
\begin{equation}
    (R^{jj_1}_{m})_{\mathbf{p}}
    =
    \sum_{j_2}
    \left(
        \frac{i}{\hbar}B_{m;m0}^{\,j; ,j_{1}\,j_{2}}
        (f_{0}^{j_{2}})_{\mathbf{p}}
        -
        \frac{i\mathbf{k}}{2}C_{m;m0}^{\,j;\,j_{1},j_{2}}
        \partial_{\bf p}(f_{0}^{j_{2}})_{\mathbf{p}}
    \right)
    \label{eq:R}
\end{equation}
Note that the coupled Eqs.~\eqref{eq:Sysf} split into $4F + 1$ subsystems, each of which corresponds to a given component $m$ and contains $2F+1-|m|$ equations. In addition, each subsystem represents the coupled linear equations with respect to index $j$ and the integral equations with respect to momentum ${\mathbf{p}}$. Therefore, as the first step, we resolve the system of linear equations with respect to $\{\tilde{f}^{j}_{m}\}_{\mathbf{p}}$ by employing the Cramer's rule:
\begin{equation}
    \{\tilde{f}^{j}_{m}\}_{\mathbf{p}}
    =
    \frac{
        \hat{\Delta}^{j}_{m}[(g^{j_1}_{m})_{\mathbf{pk}}
        -
        \sum_{j_2}
        (R^{j_1j_2}_{m})_{\mathbf{p}}
        \{\tilde{\varepsilon}_{m}^{j_{2}}\}_{\mathbf{p}}]
    }{
        (\Delta_{m})_{\mathbf{p}}
    }
    , 
    \label{eq:Sol_f}
\end{equation}
where
\begin{equation*}
     (\Delta_{m})_{\mathbf{p}}
    =
    \det\left[
        a
        \delta^{j_1j_2}
        +
        (G^{j_1j_2}_{m})_{\mathbf{p}}
    \right]
    .
\end{equation*}
The numerator of Eq.~\eqref{eq:Sol_f} represents the determinant of matrix, which is formed from $   
    a
    \delta^{j_1j_2}
    +
    (G^{j_1j_2}_{m})_{\mathbf{p}}
$ 
by replacing the $j$-th column with 
$
    (g^{j}_{m})_{\mathbf{pk}}
    -
    \sum_{j_1}
    \{\tilde{\varepsilon}_{m}^{j_{1}}\}_{\mathbf{p}}
    (R^{jj_1}_{m})_{\mathbf{p}}.
$
For the sake of brevity, in writing the results below, we present it in terms of a certain operator $\hat{\Delta}^{j}_{m}$, whose action on an arbitrary function $b^{j_{1}}_{m}$ is defined as
\begin{equation*}
    \hat{\Delta}^{j}_{m}[b^{j_{1}}_{m}]
    =
    \sum_{j_1\geq |m|}
    (-1)^{j+j_1}
    b^{j_{1}}_{m}
    \det_{j_3\neq j_1,\: j_4\neq j}
    \left[
        a
        \delta^{j_3j_4}
        +
        (G^{j_3j_4}_{m})_{\mathbf{p}}
    \right].
\end{equation*}
Substitution of the found solution \eqref{eq:Sol_f} into \eqref{eq:PertEn} with accounting for the Fourier-Laplace transform yields,
   \begin{gather}
    \{\tilde{\varepsilon}^{j}_{m}\}_{\mathbf{p}} 
    +
    \frac1V
    \sum_{{\bf p}'}
    J^{[j]}({\bf p}-{\bf p}')
    \frac{\hat{\Delta}^{j}_{m}[\sum_{j_2}
    \{\tilde{\varepsilon}_{m}^{j_{2}}\}_{\mathbf{p}'}
    (R^{j_1j_2}_{m})_{\mathbf{p}'}]}{(\Delta_{m})_{\mathbf{p}'}}
    =
    \nonumber
    \\
    \frac1V
    \sum_{{\bf p}'}
    J^{[j]}({\bf p}-{\bf p}')
    \frac{
        \hat{\Delta}^{j}_{m}[(g^{j_1}_{m})_{\mathbf{p}'\mathbf{k}}]
    }{
        (\Delta_{m})_{\mathbf{p}'}
    }
   . 
\end{gather}
This equation can be easily recast in the form useful for our subsequent analysis:
\begin{equation} 
    \{\tilde{\varepsilon}_{m}^{j}\}_{\mathbf{p}}
    =
    \frac1V \sum_{\mathbf{p}'}
    \sum_{j_2\geq |m|}
    (O^{jj_2}_m)^{-1}_{\mathbf{pp}'}
    \sum_{{\bf p}''}
    J^{[j_2]}({\bf p}'-{\bf p}'')
    \frac{
        \hat{\Delta}^{j_2}_{m}[(g^{j_1}_{m})_{\mathbf{p}''\mathbf{k}}]
    }{
        (\Delta_{m})_{\mathbf{p}''}
    }
    ,
    \label{eq:FinEq}
\end{equation}
where 
\begin{equation}
   (O^{j_{1}j_{2}}_{m})^{-1}_{\mathbf{p}'\mathbf{p}''}
    =
    \dfrac{\mathop{\rm adj}((O^{j_{1}j_{2}}_{m})_{\mathbf{p}'\mathbf{p}''})}{\det[(O^{j_{3}j_{4}}_{m})_{\mathbf{p}_{1}\mathbf{p}_{2}}]},
    \quad
     \sum_{j_{1}\geq |m|,\mathbf{p}'}(O^{jj_{1}}_{m})_{\mathbf{pp}'}(O^{j_{1}j_{2}}_{m})^{-1}_{\mathbf{p}'\mathbf{p}''}=\delta^{jj_{2}}\delta_{\mathbf{pp}''},
    \label{eq:Det}
\end{equation}
and
\begin{flalign}
    &(O^{jj_1}_m)_{\mathbf{pp}'}
    =
    \nonumber\\
    &\sum_{j_2\geq |m|}
    \left[
        \delta^{jj_1}
        \delta^{j_1j_2}
        \delta_{\mathbf{pp}'}
        +
        \vphantom{\frac1V}
        \frac{J^{[j]}({\bf p}-{\bf p}')}{V(\Delta_{m})_{\mathbf{p}'}}
    (-1)^{j+j_2}
    (R^{j_2j_1}_{m})_{\mathbf{p}'}
    \det_{j_3\neq j_2,\: j_4\neq j}
    \left[
        a
        \delta^{j_3j_4}
        +
        (G^{j_3j_4}_{m})_{\mathbf{p}'}
    \right]
    \right]
   . 
   \label{eq:O1}
\end{flalign}
Thus, the poles of the right-hand side of equation \eqref{eq:FinEq} determine all possible normal and abnormal oscillations in Bose and Fermi systems. Quantum statistics is given by the sign in the equation \eqref{eq:PartEn} -- the minus sign corresponds to fermions and plus to bosons. Remind also that the normal oscillations that we examine below do not depend on the initial condition.

\section{Dispersion equation and its solution}\label{sec:Solution}

In this section, we focus on normal oscillations, which, according to Eqs.~\eqref{eq:FinEq} and \eqref{eq:Det}, are determined by the zeros of $\det[(O^{jj_1}_m)_{\mathbf{pp}'}]$. This determinant must be simultaneously calculated in angular momentum ($j$) and momentum ($\mathbf{p}$) domains (spaces).  

\subsection{The case of zero magnetic field, \texorpdfstring{$h^j_m=0$}{hjm=0}}
\label{ssec:hjm=0}

To understand how we should treat $\det[(O^{jj_1}_m)_{\mathbf{pp}'}]$, let us explore a more simple case when at $t=0$ the external field is turned off. Then among all components of the Wigner equilibrium distribution function only $(f^0_0)_{\mathbf{p}}\neq 0$.
Due to this fact, we can simplify the functions $(G^{jj_1}_{m})_{\mathbf{p}}$ and $(R^{jj_1}_{m})_{\mathbf{p}}$ by employing Eqs.~\eqref{eq:BCCoeff}:
\begin{equation*}
    (G^{jj_1}_{m})_{\mathbf{p}}
    =
    \frac{
        i\mathbf{k}
        \partial_{\mathbf{p}}
        (\varepsilon_{0}^{0})_{\mathbf{p}}
    }{
        \sqrt{2F+1}
    }
    \delta^{jj_1}
    ,
    \quad
    (R^{jj_1}_{m})_{\mathbf{p}}
    =
    -
    \frac{
        i\mathbf{k}
        \partial_{\mathbf{p}}
        (f_{0}^{0})_{\mathbf{p}}
    }{
        \sqrt{2F+1}
    }
    \delta^{jj_1}
    .
\end{equation*}
Hence, following Eq.~\eqref{eq:Sysf}, we immediately conclude that all components of the Wigner distribution function oscillate independently and operator $(O^{jj_1} _m)_{\mathbf{pp}'}$ takes a diagonal form in the $j$-domain,   
\begin{equation*}
    (O^{jj_1} _m)_{\mathbf{pp}'}
    \equiv
    (O^{j})_{\mathbf{pp}'}\delta^{jj_1}
    , 
\end{equation*}
where $(O^{j})_{\mathbf{pp}'}$ is found from Eq.~\eqref{eq:O1},
\begin{equation}
    (O^{j})_{\mathbf{pp}'}
    =
        \delta_{\mathbf{pp}'}
        -
        \frac1V
        J^{[j]}({\bf p}-{\bf p}')
    \frac{
        i\mathbf{k}
        \partial_{\mathbf{p}'}
        (f_{0}^{0})_{\mathbf{p}'}
    }{
        a\sqrt{2F+1}+i\mathbf{k}
        \partial_{\mathbf{p}'}
        (\varepsilon_{0}^{0})_{\mathbf{p}'}
    }
   .
\end{equation}
Thus, all that remains is to calculate the determinant $\det[(O^{j})_{\mathbf{pp}'}]$. 
Note that the respective matrix has an interesting property.  
For fully or partially degenerate gases, it resembles the identity matrix, except for a vertical "{}stripe"{} the width of which is given by $i\mathbf{k}
        \partial_{\mathbf{p}'}
        (f_{0}^{0})_{\mathbf{p}'}$.       
This matrix structure allows us to reduce the computation of the original determinant to solving another determinant corresponding to a certain truncated matrix with diagonal being a part of the main diagonal of the original matrix. The mentioned stripe contains the truncated matrix whose order coincides with the stripe width. Therefore, the diagonal element of the $n$-column of the truncated matrix is written in the form:
\begin{equation*}
    1 
    -
    \frac1V
    J^{[j]}(0)
    \frac{
        i\mathbf{k}
        \partial_{\mathbf{p}'_{n}}
        (f_{0}^{0})_{\mathbf{p}'_{n}}
    }{
        a\sqrt{2F+1}+i\mathbf{k}
        \partial_{\mathbf{p}'_{n}}
        (\varepsilon_{0}^{0})_{\mathbf{p}'_{n}}
    }.
\end{equation*}
Finally, when computing the determinant of the truncated matrix, we should keep only the linear terms in interaction which emerge from the entries of its main diagonal. This is because the kinetic equation itself is of the first order in interaction in the collisionless approximation. 
Therefore, we arrive at the following dispersion equation:
\begin{equation}
    1 
    -
    \frac{1}{V}\frac{J^{[j]}(0)}{\sqrt{2F+1}}
    \sum_{\mathbf{p}}
    \frac{
        i\mathbf{k}
        \partial_{\mathbf{p}}
        (f_{0}^{0})_{\mathbf{p}}
    }{
        a
        +
        (i\mathbf{k}
        \mathbf{p}/M)
    }
    =
    0
    .
    \label{eq:DispEq}
\end{equation}
Note that as the temperature increases, the width of the stripe  enlarges and in the limit of a classical gas, it coincides with the entire matrix $(O^{j})_{\mathbf{pp}'}$. In addition, in the linear order in interaction, the computation of the determinants of the truncated and original matrices produces the same result. Nevertheless, working with the determinant of a truncated matrix significantly simplifies the calculations in the case of higher order terms in interaction. 

Now we find the excitation spectra of fully degenerare quantum gases by using the dispersion equation \eqref{eq:DispEq}.

For a \textit{Bose gas} at zero temperature, the distribution function with discrete momentum is given by $f^{[\alpha]}(\mathbf{p})=n_{0}V\delta_{\mathbf{p}0}$, 
where $n_{0}$ is the condensate density. However, when converting to integration (continuous momentum), we should make a replacement: $f^{[\alpha]}(\mathbf{p})\to n_{0}(2\pi\hbar)^{3}\delta(\mathbf{p})$. Therefore, following Eq.~\eqref{eq:EqWig}, in which $(T^{0}_{0})_{\alpha\beta}=(1/\sqrt{2F+1})\delta_{\alpha\beta}$, one obtains
\begin{equation}
    (f^{0}_{0})_{\mathbf{p}}
    =
    \sqrt{2F+1}
    (2\pi\hbar)^{3}
    n_{0}
    \delta({\bf p})
    .
    \label{eq:fBose}
\end{equation}
Finally, replacing in Eq.~\eqref{eq:DispEq} the sum with integral and performing integration with the given $(f^{0}_{0})_{\mathbf{p}}$, we have
\begin{equation}
    \omega^{2}
    =
    \frac{n_{0}J^{[j]}(0)}{M}
    k^{2}
    , 
\end{equation}
where $\omega=ia$ is the oscillation frequency. The interaction parameter determined by Eq.~\eqref{eq:PartEn} should be chosen according to Bose statistics. It is clear that we are dealing with $2F+1$ ($j=0\dots 2F$) undamped modes. The obtained result agrees with the celebrated Bogoliubov spectrum \cite{Bogoliubov1947} at small wave vectors (phonon spectrum). It should be noted that we do not obtain here the exact Bogoliubov result because we assumed above that the spatial inhomogeneities are small (large wavelength). This corresponds to a small wave vector $\mathbf{k}$. 

For a {\it ground-state Fermi gas}, the tensorial component $(f^{0}_{0})_{\mathbf p}$ of the Wigner function is written in terms of the Heaviside step function $\Theta(\varepsilon)$,
\begin{equation}
    (f^{0}_{0})_{\mathbf{p}}
    =
    \sqrt{2F+1}
    \Theta(\varepsilon_{\mathrm{F}}-\varepsilon_{\mathbf{p}})
    ,
\end{equation}
where $n$ is the particle density and $p_{\mathrm{F}}$ is the Fermi momentum. Therefore, by computing the respective integral in Eq.~\eqref{eq:DispEq}, we have the following equation:
\begin{equation}
    \frac{\xi}{2}
    \ln\frac{\xi+1}{\xi-1}
    -
    1
    =
    \frac{
        2\pi^2\hbar^3
    }{
        J^{[j]}(0)Mp_{\mathrm{F}}
    } 
    \label{eq:ZeroSound}
\end{equation}
which governs the the quantity
\begin{equation*}   
    \xi
    =
    \frac{aM}{ikp_{\rm F}}.
\end{equation*}
Just as for bosons, we have $2F+1$ oscillation modes. These oscillations are well known as zero sound and were predicted by Landau in the framework of the Fermi liquid theory \cite{Landau_1957}. The zero sound dispersion law may indicate the violation of the Pomeranchuk stability conditions \cite{Pomeranchuk_1958} (the normal state of a Fermi liquid becomes unstable) which signals a phase transition to another state \cite{Peletminskii_LTP_II_1999}. In the context of quantum gases, zero sound oscillations have been studied for dipolar \cite{Bohn_2010,Shlyapnikov_2013} and spinor \cite{Yip_PRA_1999,Bulakhov_JPhysA_2023} gases.

Remind that in general case, the Laplace parameter is complex and, consequently, Eq.~\eqref{eq:ZeroSound} determines both oscillation frequencies and damping factor. Namely, we are dealing with two solutions: undamped and damped one, which are also known as "fast"{} and "slow"{} waves, respectively \cite{Poluektov_2014}.

\subsection{The case of nonzero magnetic field, \texorpdfstring{$h^j_m\neq0$}{hjm=0}}

Now we can proceed to calculating the determinant and solving the dispersion equation in the presence of the external field $h^{j}_{m}=h^{j}_{0}\delta_{m0}$.

Before addressing the structure of the matrix $(O^{jj_1}_m)_{\mathbf{pp}'}$, note that we are considering an ideal degenerate gas of atoms, as in subsection \ref{ssec:hjm=0}. Since the external magnetic field removes the spin degeneracy, the structure of the equilibrium state of a degenerate gas becomes more complicated and essentially depends on statistics. In a Fermi gas, several Fermi energies occur, one for each spin projection. In a ground-state ideal Bose gas, on the contrary, the state is specified by only one spin projection taking the maximum value (Bose-Einstein condensate in ferromagnetic state \cite{Akhiezer_JETP_1998}). 

Taking into account that $j$-components of $\{\tilde{f}^{j_{1}}_{m}\}_{\mathbf{p}}$ with the given $m$ are mixed up (see Eq.~\eqref{eq:Sysf}) we can conclude that the structure of the matrix $(O^{jj_1}_m)_{\mathbf{pp}'}$ is similar to that described in the previous subsection but with a larger number of stripes: $(2F+1)(2F+1-|m|)$ for fermions and $(2F+1-|m|)$ for bosons. Thus, the determinant of interest is reduced to the product of determinants of low-order matrices, the number of which is defined by statistics. These low-order matrices belong to different stripes (one per stripe) and have the same structure as the truncated matrix from subsection \ref{ssec:hjm=0}.

In the linear order in interaction, we come to following dispersion equation:
\begin{gather}
    1
    +
    \sum_{j,j_{1}\geq |m|}
    \frac{J^{[j]}(0)}{V}
    \sum_{\mathbf{p}}
    \frac{(-1)^{j+j_1}}{(\Delta_{m})_{\mathbf{p}}}
    (R^{j_1j}_{m})_{\mathbf{p}}
    \det_{j_3\neq j_1,\: j_4\neq j}
    \left[
        a
        \delta^{j_3j_4}
        +
        (G^{j_3j_4}_{m})_{\mathbf{p}}
    \right]
    =
    0
    .
   \label{eq:DispEqh}
\end{gather}
From this moment, we treat   
$(G^{j_3j_4}_{m})_{\mathbf{p}}$ as the quantity in the zeroth order in interaction,
\begin{equation*}
    (G^{j_{3}j_4}_{m})_{\mathbf{p}} 
    \approx
    \frac{i\mathbf{k}\mathbf{p}}{M}
    \delta^{j_{3}j_4}
    +
    \sum_{j_2}
    \frac{i}{\hbar}
    B_{m;m0}^{\,j_{3}; ,j_{4}\,j_{2}}
    h_{0}^{j_{2}}
    .
\end{equation*}
Equation \eqref{eq:DispEqh} solves (in quadratures) the declared problem of finding the high-frequency excitation spectrum for a gas of atoms with arbitrary spin and non-contact interaction in the presence of an external field. In contrast to previous studies \cite{Yip_PRA_1999}, we do not parameterize the interaction by the $s$-wave scattering length. The corresponding parameter $J^{[j]}(0)$ appears in a natural way. Moreover, for integrating the dispersion equation, there is no necessity to require the independence of interaction function on angle between two momenta (see, e.g., \cite{Landau_1957}). 

Since in the general case solving Eq.~\eqref{eq:DispEqh} represents a cumbersome computational problem, we limit ourselves below to studying modes with $|m|=2F$. Then, according to Eq.~\eqref{eq:DispEqh}, the term with $j=2F$ remains from the entire sum, and the entering determinants are significantly simplified. To specify the unperturbed state, we consider, as above, degenerate Fermi and Bose gases with arbitrary spin $F$ in equilibrium.

For a {\it ferromagnetic Bose gas} in an external field, the generalization of Eq.~\eqref{eq:fBose} gives 

\begin{equation}
    f^{[\alpha]}(\mathbf{p})
    =
    \frac{
        2\pi^2\hbar^{3}n_{0}
    }{
        M\sqrt{2M\varepsilon_{\mathbf{p}}}
    }
    \delta_{\alpha 1}
    \delta(
        \varepsilon_{\mathbf{p}}
        -
       \varepsilon_{\mathrm{B}}
    )
    , \quad  
     \varepsilon_{\mathrm{B}}=
     \sum_{j}h^{j}_{0}(T^{j}_{0})_{11}.
    \label{eq:FerrCond}
\end{equation}
When replacing the summation over ${\mathbf{p}}$ by integration with the function given by \eqref{eq:FerrCond}, we obtain the total density equal to the condensate density $n_{0}$ of atoms with the maximum spin projection. Furthermore, we assume that the element $(T^{j}_{0})_{11}$ corresponds to the largest projection of the spin operator $F_{z}$. Indeed, although the operator ${T}^{j}_{0}$ is a $j$-degree polynomial of $F_z$, we are free to arrange the spin projections on the diagonals of $T^{j}_{0}$ -- from larger to smaller one (as we do) or vice versa, see \ref{app:STO}. 
Hence, following Eq.~\eqref{eq:EqWig}, we have
\begin{equation}
    (f^{j}_{0})_{\mathbf{p}}
    =
    \frac{
        2\pi^2\hbar^{3}n_{0}
    }{
        M\sqrt{2M\varepsilon_{\mathbf{p}}}
    }
    (T^{j}_{0})_{11}
    \delta(
        \varepsilon_{\mathbf{p}}
        -
        \varepsilon_{\mathrm{B}}
    ).
\end{equation}
This function determines the quantity $(R^{j_{1}j}_{m})_{\mathbf{p}}$ entering the dispersion equation. Next, the straightforward integration of Eq.~\eqref{eq:DispEqh} yields, 
\begin{gather}
    1
    +
    n_0
    J^{[j]}(0)
    \sum_{j_2}
    (T^{j_2}_{0})_{11}
    \left(
        \frac{
            \sqrt{M}B^{j_2}
        }{
            2^{3/2}
            \hbar
            \sqrt{\varepsilon_{\mathrm{B}}}
            k
        }
        \ln
        \frac{w+k}{w-k}
        -
        \frac{
            C^{j_2}
        }{
            4
            \varepsilon_{\mathrm{B}}
        }
        \frac{k^{2}}{w^2-k^{2}}
    \right)
    \Theta(\varepsilon_{\mathrm{B}})
    =
    0,
    \label{eq:DispBoseH}
\end{gather}
where
\begin{equation*}   
    w=
    \frac{
        \sqrt{M}
    }{
        i\sqrt{2\varepsilon_{\mathrm{B}}}
    }
    \left(
        a
        +
        \frac{i}{\hbar}
        \sum_{j_{2}}
        B^{j_2}
        h^{j_2}_{0}
    \right)
\end{equation*}
and $B^{j_2}=B^{\ 2F;\ 2F\,j_2}_{\pm2F;\pm2F 0}$, $C^{j_2}=C^{\ 2F;\ 2F\,j_2}_{\pm2F;\pm2F0}$. Applying the method of successive approximation to Eq.~\eqref{eq:DispBoseH}, we obtain a series for $\omega=ia$ up to $k^2$:
\begin{equation}\label{eq:SpecBose}
    \omega
    \approx
    \omega_0
    +
    \omega_2
    k^2,
\end{equation}
where
\begin{equation*}
    \omega_0
    =
    \frac{1}{\hbar}
    \sum_{j_{2}}
    B^{j_2}
    \left(
        h^{j_2}_{0}
        +
        n_0
        J^{[j]}(0)
        (T^{j_2}_{0})_{11}
        \Theta(\varepsilon_{\mathrm{B}})
    \right) 
\end{equation*}
and
\begin{equation*}
    \omega_2
    =
    \hbar
    \frac{
        \left(
            4\varepsilon_B
            +
            3n_0
            J^{[j]}(0)
            \sum_{j_2}
            C^{j_2}
            (T^{j_2}_{0})_{11} \Theta(\varepsilon_{\mathrm{B}})
        \right)
    }{  
        6M
        n_0
        J^{[j]}(0)
        \sum_{j_2}
        B^{j_2}
        (T^{j_2}_{0})_{11} \Theta(\varepsilon_{\mathrm{B}})
    }.
\end{equation*}
It is worth stressing that this result and rewsults below are valid for $\hbar^{2}k^{2}/2M\ll n_{0}J^{[j]}(0)$ \cite{klimontovich_JETP_1952}. The gapful mode, given by Eq.~\eqref{eq:SpecBose}, corresponds to ferromagnetic spin excitations. Its structure is similar to that obtained for the ferromagnetic state of a weakly interacting spin-1 gas with condensate \cite{Ohmi_JPSJ_1998,Ho_PRL_1998,Akhiezer_JETP_1998,Peletminskii_PhysA_2007,Ueda_PhysRep_2012,Bulakhov_JPhysA_2021}. However, the difference is due to the fact that the indicated works use the non-zero chemical potential to calculate the single-particle excitations.   

For a {\it Fermi gas} at zero temperature, we have 
$
    f^{[\alpha]}(\mathbf{p})
    =
    \Theta
    (
        \varepsilon_{\rm F}^{[\alpha]}
        -
        \varepsilon_{\bf p}
    )
    .
$
Consequently, following again Eq.~\eqref{eq:EqWig}, one finds  
\begin{equation}\label{eq:DFZero}
    (f^{j}_{0})_{\mathbf{p}}
    =
    \Theta
    (
        \varepsilon_{\rm F}^{[\alpha]}
        -
        \varepsilon_{\bf p}
    )
    (T^{j}_{0})_{\alpha\alpha}
    ,
\end{equation}
where 
\begin{equation*}\label{eq:FermiEn}
    \varepsilon_{\mathrm{F}}^{[\alpha]}
    =
    \varepsilon_{\rm F}(h^{j}_{0})
    +
    \sum_{j_{1}=0}
    h^{j_1}_{0}
    (T^{j_1}_{0})_{[\alpha\alpha]}
    .
\end{equation*}
Remind that no summation is assumed over the indices in square brackets. Calculation of the total number of spin-$F$ atoms yields
\begin{equation} \label{eq:EnF}
    \sum_{\alpha=1}^{2F+1}
    \Theta(\varepsilon_{\rm F}^{[\alpha]})
    [\varepsilon_{\rm F}^{[\alpha]}]^{3/2}
    =
    (2F+1)
    [\varepsilon_{\rm F}(0)]^{3/2}
    , 
    \quad
    \varepsilon_{\rm F}(0)=\dfrac{\left(3\pi^2n\right)^{2/3}\hbar^2}{2^{1/3}(2F+1)^{2/3}M}
    ,
\end{equation}
where $n$ is the total atomic density. Next, integration of Eq.~\eqref{eq:DispEqh} results in
\begin{gather}
    1
    +
    J^{[j]}(0)
    \sum_{j_2}
    (T^{j_2}_{0})_{\alpha\alpha}
    \left(
        \frac{iB^{j_2}}{(2\pi\hbar)^3\hbar}
        \int
        \mathop{d^3p}
        \frac{
            \Theta
            (
                \varepsilon_{\rm F}^{[\alpha]}
                -
                \varepsilon_{\bf p}
            )
        }{
            i\frac{\mathbf{kp}}{M}
            +
            \frac{i}{\hbar}
            \sum_{j_{2}=0}
            B^{j_2}
            h^{j_2}_{0}
            +
            a
        }
        +
    \right.
        \nonumber
        \\
    \left.
        \frac{
            \sqrt{\varepsilon_{\mathrm{F}}^{[\alpha]}}
            M^{3/2}
            C^{j_2}
            \Theta(\varepsilon^{[\alpha]}_{\mathrm{F}})
        }{
            2^{3/2}
            \pi^2
            \hbar^3
        }
        \left[
            1
            -
            \frac{w^{[\alpha]}}{2k}
            \ln\frac{w^{[\alpha]}+k}{w^{[\alpha]}-k}
        \right]
    \right)=0
    ,
\end{gather}
where
\begin{equation*}   
    w^{[\alpha]}=
    \frac{
        \sqrt{M}
    }{
        i\sqrt{2\varepsilon^{[\alpha]}_{\mathrm{F}}}
    }
    \left(
        a
        +
        \frac{i}{\hbar}
        \sum_{j_{2}=0}
        B^{j_2}
        h^{j_2}_{0}
    \right)
    .
\end{equation*}
We represent the second term in quadratures (through the integral) because it cannot be treated using the residue theorem due to the non-analyticity of the Heaviside step function. However, since we are interested in the long-wave limit (small values of $k$), this does not prevent us, as before, to apply the method of successive approximations. As a result, we obtain a series for $\omega=ia$ up to $k^{2}$:
\begin{equation}
    \omega
    \approx
    \omega_0
    +
    \omega_2
    k^2,
    \label{eq:FerSpinWave}
\end{equation}
where
\begin{equation*}
    \omega_{0}
    =
    \frac{1}{\hbar}
    \sum_{j_{2}}B^{j_{2}}
    \left(
        h_{0}^{j_{2}}
        +
        \frac{nJ^{[j]}(0)}{2F+1}
        (T_{0}^{j_{2}})_{\alpha\alpha}
        \left(
            \frac{
                \varepsilon_{\rm F}^{[\alpha]}
            }{
                \varepsilon_{\rm F}(0)
            } 
        \right)^{3/2}
        \Theta(\varepsilon^{[\alpha]}_{\mathrm{F}})
    \right)
\end{equation*}
and
\begin{flalign*}
    &\omega_{2}
    =
    \\
    &
    \frac{\hbar}{2M}
    \frac{
        \sum_{j_{1}}
        C^{j_{1}}
        (T_{0}^{j_{1}})_{\alpha\alpha}(\varepsilon_{\rm F}^{[\alpha]})^{3/2}
        \Theta(\varepsilon^{[\alpha]}_{\mathrm{F}})
    }{
        \sum_{j_{2}}
        B^{j_{2}}
        (T_{0}^{j_{2}})_{\beta\beta}(\varepsilon_{\rm F}^{[\beta]})^{3/2}
        \Theta(\varepsilon^{[\beta]}_{\mathrm{F}})
    }
    +
    \frac{2\hbar(2F+1)(\varepsilon_{\rm F}(0))^{3/2}}{5MnJ^{[j]}(0)}
    \frac{
        \sum_{j_{1}}
        B^{j_{1}}
        (T_{0}^{j_{1}})_{\alpha\alpha}(\varepsilon_{\rm F}^{[\alpha]})^{5/2}
        \Theta(\varepsilon^{[\alpha]}_{\mathrm{F}})
    }{
        \left(
            \sum_{j_{2}}
            B^{j_{2}}
            (T_{0}^{j_{2}})_{\beta\beta}(\varepsilon_{\rm F}^{[\beta]})^{3/2}
            \Theta(\varepsilon^{[\beta]}_{\mathrm{F}})
        \right)^{2}
    }
    .
\end{flalign*}
This gapful undamped mode corresponds to paramagnetic spin excitations. Note that there exist another damped wave. However, we do not provide explicit expressions for the real and imaginary parts of the corresponding frequency due to their cumbersomeness. The kinetic theory of spin waves (for $F=1/2$) was also developed within the framework of the Fermi liquid approach \cite{Silin_1958_sw}. 
Note that for spin-1/2 atoms in a weak magnetic field, equation \eqref{eq:FerSpinWave} reproduces the frequencies of spin waves obtained in Ref.~\cite{Yip_PRA_1999}.

Thus, having obtained partial solutions corresponding to $|m|=2F$, we showed the possibility of propagation of spin waves both for bosonic and fermionic atomic gases. The complete spectrum of oscillation modes corresponding to other values of $m$ is found by solving the general dispersion equation \eqref{eq:DispEqh} for a particular spin value. However, some additional conclusions can be drawn for an arbitrary spin. In general, the physical nature of collective modes is governed by the tensorial components of the Wigner density matrix. For a given $m$, the oscillations are characterized by the Wigner distribution functions of rank $j=|m|\dots 2F$. Oscillations with $m\neq 0$ are characterized by a quadratic dispersion law (with a gap) and correspond to spin waves. As for the excitations with $m=0$, they are determined by the components of the tensorial Wigner distribution function of all ranks. The respective modes have a linear dispersion law and represent zero sound or density excitations. For fermions, as we discuss below, the existence of zero sound modes is especially sensitive to the dependence of the equilibrium configuration on the magnetic field.

\section{Results and discussion}\label{sec:Discussion}

Employing the general principles of angular momentum theory and apparatus of irreducible spherical tensor operators, we have formulated a Hamiltonian for a system of identical particles with arbitrary spin $F$. It accounts for their coupling with external field and pairwise interaction through spin-induced multipolar moments. As the next step, we have derived the collisionless kinetic equation for quantum gases valid for small inhomogeneities.   

Before proceeding with the linearizing kinetic equation, we have assumed that the interaction potential has a sharp peak so that everywhere below the mean-field particle energy was  given by Eq.~\eqref{eq:QuasEnXSimp}. This approximation is indeed suitable when dealing with gases of neutral atoms but invalidates in the case of strong Coulomb interaction, as arises in plasma. However, by conducting a minor analysis of the general formula for the particle energy in terms of general pair potential \cite{AkhPel,Bulakhov_JPhysA_2023}, one can conclude that the replacement $J^{[j]}(0)\to J(k)$ ($J(k)\propto 1/k^2$) in the dispersion equation \eqref{eq:DispEq} is sufficient to describe the normal oscillations in plasma  \cite{Akhiezer1975}.

Having obtained equation \eqref{eq:FinEq}  describing all possible oscillations, we have analyzed the case of zero external field to reproduce the well-known results, such as the long-wave limit of the Bogoliubov spectrum \cite{Bogoliubov1947} and Landau zero sound oscillations \cite{Landau_1957}. In addition, this case allows us to understand how to treat the determinant that defines the dispersion equation. In particular, we have showed that there is no need to make any assumptions about the form of the pairwise interaction potential $J^{[j]}({\mathbf{p}})$ when calculating the integrals in the dispersion equation since the later contains the quantity $J^{[j]}(0)$, which appears in a natural way. This result also extends to the case of non-zero external field.

For a gas of fermionic atoms in the polarized equilibrium state \eqref{eq:DFZero}, the dispersion equation \eqref{eq:DispEqh} for zero sound ($m=0$) loses its meaning since all the interaction terms are canceled.
This fact can be proved at least for $F=1/2,\,3/2,\,5/2$ by considering  Eq.~\eqref{eq:PartEn} and using the properties of $(R_{0}^{jj_{1}})_{\mathbf{p}}$ and $A^{j'j}$ (see Eqs.~\eqref{eq:R}, \eqref{eq:APropA}).  
We believe that this conclusion is valid for an arbitrary half-integer spin. 
It is worth noting that by choosing another equilibrium state \cite{Silin_1958_sw,Yip_PRA_1999} (weak magnetic field) these zero sound modes are restored. If we assume that $J^{[j]}(0)$ corresponds to contact interaction, then the result of no zero sound agrees with that for spin-3/2 atoms obtained in a slightly different formulation \cite{Bulakhov_JPhysA_2023} (employing the spherical tensor operators). Systems of spin-3/2 atoms have recently attracted considerable interest \cite{Lewenstein_PRL_2013, Kosmachev_JETP_2015,  Bulakhov_JPhysA_2023,Kovalevsky_AnnPhys_2014,Xu_AnnPhys_2021,Xu_JPhysB_2022}.

As for the physical relevance of multipolar exchange interaction, it can be proved that the quadrupolar interaction is related to the dipole-dipole forces \cite{Racah_1959}. Therefore, for atoms with large intrinsic dipole moments, such as Erbium \cite{Erbium_PRL_2012,Erbium_PRA_2022} and Dysprosium \cite{Dysprosium_PRL_2011,Dysprosium_PRA_2015}, the quadrupolar exchange interaction is essential \cite{PitStr2016,Bulakhov_JPhysA_2022}. 
Apparently, octupolar and higher order interactions are weak, however, the discovery of effects, where they could manifest themselves, deserves special attention. 
In particular, the measurement of collective oscillation frequencies could potentially provide insight into the character and symmetry of interaction.

In experiments, ultracold gases are placed in magnetic or magneto-optical traps with significantly non-uniform field. Therefore, a natural question arises as for the relevance of the developed approach to realistic situation. In this regard, we note that following \cite{Lewenstein_PRL_2013}, the proposed approach can be generalized for trapping field. Nevertheless, in its current form, it allows one to employ the general equations \eqref{eq:FinEq}, \eqref{eq:DispEqh} and to get a fairly complete understanding of the structure and nature of high-frequency collective excitation spectra for a particular atomic spin. Moreover, a distinctive feature of the developed approach is its universality: suitability for arbitrary spin and both quantum statistics as well as complete and consistent consideration of multipolar degrees of freedom.   

\section*{Acknowledgements}
 The authors acknowledge support by the National Research Foundation of Ukraine, Grant No. 0124U004372. M. Bulakhov acknowledges support by STCU project “Magnetism in Ukraine Initiative”, Grant No. 9918.

\begin{appendix}

\section{Matrix equivalents of spherical tensor operator}

Although the theory of irreducible spherical tensor operators is widely discussed in the literature \cite{Wigner_1959,Racah_1959,Brink_1968,Sakurai_2020}, we briefly present the main formulas used in the paper. The matrix elements of the spherical tensor operator $\left(T^j_m\right)_{\alpha\beta}$ (the upper and lower indices denote its rank and component, respectively) are given by
\label{app:STO}
\begin{equation}
    \left(T^j_m\right)_{\alpha\beta}
    =
    (-1)^{\alpha-1}
    \left(
    \begin{array}{w{c}{5em}w{c}{1.6em}w{c}{5em}}
        F           &   j   & F\\
        \alpha-F-1  &   m   & F+1-\beta
    \end{array}
    \right)
    \left<F||T^j||F\right>
    ,
    \label{eq:Equivalents}
\end{equation}
where $\alpha,\beta=1\dots 2F+1$, the 2 × 3 array represents the 3-j Wigner symbol, $\left<F||T^j||F\right>=\sqrt{2j+1}$ is the Racah's normalization and $F$ is the spin of a particle.

Properties of matrix equivalents of the spherical tensor operators:
\begin{itemize}
    \item Normalization condition
\begin{equation}\label{eq:Norm}
\Tr T_{m}^{j}T^{j'}_{-m'}=(-1)^{m}\delta_{jj'}\delta_{m,m'}
.
\end{equation}
\item Fierz identity
\begin{equation}
    \sum_{j=0}^{2F}
    \sum_{m=-j}^{j}
    (-1)^m(T_{m}^{j})_{\alpha\beta}(T^{j}_{-m})_{\gamma\sigma}
    =
    \delta_{\alpha\sigma}
    \delta_{\gamma\beta}
    .
    \label{eq:FierzId}
\end{equation}

\item Selection rule
\begin{equation}
\begin{gathered}
        (T_{m_1}^{j_1})_{\alpha\beta}
    (T_{m_2}^{j_2})_{\beta\gamma}
    =
    \sum_{j=|j_1-j_2|}^{j_1+j_2}
    B(j_1,j_2,m_1,m_2;j,k)
    (T^j_k)_{\alpha\gamma}
    , \\
    k=m_1+m_2
    ,\ 
    |k|\leq j\leq 2F. \label{eq:SelRule}
\end{gathered}
\end{equation}
\item Fierz-like identity
\begin{equation} \label{eq:Fierz-like}
    \sum_{m=-j}^{j}
    (-1)^m
    (T_{m}^{j})_{\alpha\beta}
    (T_{m'}^{j'})_{\beta\gamma}
    (T^{j}_{-m})_{\gamma\sigma}
    =
    A^{jj'}
    (T_{m'}^{j'})_{\alpha\sigma},
\end{equation}
\begin{proof}
Suppose the identity does not hold. Then the following decomposition must be true:
\begin{equation}
    \sum_{m=-j}^{j}
    (-1)^m
    (T_{m}^{j})_{\alpha\beta}
    (T_{m'}^{j'})_{\beta\gamma}
    (T^{j}_{-m})_{\gamma\sigma}
    =
    A^{jj'}
    (T_{m'}^{j'})_{\alpha\sigma}
    +
    \sum_{j_1\neq j' \land m_1\neq m'}
    B^{j_1j'}_{m_1m'}
    (T_{m_1}^{j_1})_{\alpha\sigma}
    .
    \label{eq:AProof1}
\end{equation}
From the selection rule \eqref{eq:SelRule}, it is immediately clear that $m_1=m'$. Therefore, to find $B^{j_1j'}_{m_1m'}$ we have to multiply \eqref{eq:AProof1} by $(T_{-m'}^{j_2})_{\sigma\rho}$ with $j_2\neq j'$ and take trace under normalization condition. This yields:
\begin{equation}
    \sum_{m=-j}^{j}
    (-1)^m
    (T_{m}^{j})_{\alpha\beta}
    (T_{m'}^{j'})_{\beta\gamma}
    (T^{j}_{-m})_{\gamma\sigma}
    (T_{-m'}^{j_2})_{\sigma\alpha}
    =
    B^{j_2j'}_{m'm'}
    . 
    \label{eq:AProof2}
\end{equation}
Through the irreducibility and the selection rule, the LHS of \eqref{eq:AProof2} is not equal to zero only if $j'=j_2$. This contradicts our initial assumption. Thus, \eqref{eq:Fierz-like} holds.
\end{proof}
\end{itemize}

Finally, employing Fierz identity \eqref{eq:FierzId}, one can prove  that matrix $A^{jj'}$ has non-obvious important properties:
\begin{equation}
    \sum_{j=0}^{2F}
    A^{jj'}
    =
    (2F+1)\delta^{0j'}
    ,\quad
    \sum_{j'=0}^{2F}
    A^{jj'}
    =
    \mathbf{1}^{j0}
    ,
    \label{eq:APropA}
\end{equation}
where $\mathbf{1}^{j0}$ is a column vector containing only ones.

\end{appendix}

\bibliographystyle{apsrev}
\bibliography{main}

\end{document}